\documentclass{llncs}

\usepackage{amssymb}
\usepackage{graphicx}
\usepackage{hyperref}

\usepackage{listings}
\usepackage[dvipsnames]{xcolor}
\usepackage{courier}

\newcommand{\f}[1]{\mbox{\sf #1}} 
 
\newcommand{\fs}[1]{\mbox{\scriptsize\sf #1}} 
\newcommand{\ft}[1]{\mbox{\tiny\sf #1}} 

\newcommand\COO{$\mbox{\sf CO}_{\mbox{\scriptsize\sf 2}}$}

\author{Armin Wolf}

\institute{IT4Energy Center, Fraunhofer FOKUS\\
	Kaiserin-Augusta-Allee 31, D-10589 Berlin\\
	\email{armin.wolf@fokus.fraunhofer.de}\\
	\includegraphics[height=8pt]{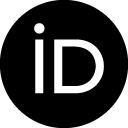}%
	~\href{https://orcid.org/0000-0003-3940-0792}{https://orcid.org/0000-0003-3940-0792}
}

\title{Modular Modeling and Optimized Scheduling\\
       of Building Energy Systems Based on\\
			 Mixed Integer Programming\thanks{The presented work was funded by the German Federal Ministry 
			for Economic Affairs and Energy within the project ``WaveSave'' (BMWi, funding number 03ET1312A).}}

\begin{document}

\lstset{language=XML,morecomment=[s]{!--}{--}}

\maketitle

\begin{abstract}
Almost climate neutral buildings are one of the core goals in terms of sustainability. Beside the support 
of the necessary design decisions for an integrated, interoperable, ecological and economical operation of building 
energy systems, innovative management solutions for scheduling the operation of decentralized energy systems are of
great importance. The challenge is optimal interaction between energy system components in terms of own consumption, 
energy efficiency and resource consumption as well as greenhouse gas emissions. To achieve these goals a modular
optimization approach based on Mixed Integer Programming is proposed. In detail, and to our knowledge the first
time, a MIP model for the dynamic behavior of fuel cell Combined Heat and Power plants is presented. Our approach 
is evaluated for the operation of heat pumps showing that their energy efficiency can be increased significantly.
\end{abstract}

\keywords{building energy systems, constraint-based scheduling and optimization,  cross-sector coupling,
energy efficiency,  Mixed Integer Programming} 

\section{Introduction}
 
In the context of the energy transformation, known as the ``Energiewende'', and global warming almost climate neutral 
buildings are one of the core goals. Beside the support of the necessary design decisions for an integrated, 
interoperable, ecological and economical operation of building energy systems, innovative management solutions for 
scheduling the operation of decentralized energy systems are of great importance. The challenge is the optimal 
interaction between energy system components in terms of own consumption, energy efficiency and resource consumption as 
well as greenhouse gas emissions. To achieve these goals, a Mixed Integer Programming (MIP) based optimization tool for 
the combination of energy system components like Combined Heat and Power plants (CHPs) and the operation of integrated 
energy systems was developed, using real or virtual costs in an overall objective function and taking into account the 
uncertainties caused by weather, volatility of renewable energies as well as the behavior and spontaneity of residents. 
This tool prototype was implemented and evaluated using previously defined application scenarios.

The paper is organized as follows. In the next section some related work is presented. In  Section~\ref{sec:optModel} 
a modular modeling approach for optimal operation of building energy systems based on MIP is presented. In detail, 
the linear modeling of the dynamic behavior of fuel cell CHPs (fcCHPs) is shown using some specific ``modeling tricks''. 
Then, in Section~\ref{sec:MIP-Optimizer} some implementation issues of our ``MIP Optimizer'' are given. Finally, in 
Section~\ref{sec:Evaluation} our approach is successfully evaluated: It is shown how the operation of heat pumps can 
be optimized significantly while reducing electric energy demand and cost without loss of comfort. The last section
concludes and points to some future work to be done.  

\section{Related Work}\label{sec:RW}

For scheduling and optimization of decentralized energy systems MIP is an adequate 
approach~\cite{bosmanPlanningProductionFleet2012,bozchaluiOptimalOperationCommercial2012,%
brahmanOptimalElectricalThermal2015} which we also use when modeling building energy systems. 
In~\cite{bosmanPlanningProductionFleet2012} microCHPs based on combustion engines are used in two different scheduling
scenarios: In the Single House Planning Problem (SHPP), the focus is on satisfying the heat demand of residents. 
The second scenario combines many microCHPs into a Fleet Planning Problem (FPP) in order to satisfy some electric
power demands, still considering domestic heat demands. Both problems are modeled as MIP problems. There, the MIP model 
of the microCHPs is rather simple compared to our MIP model for fcCHPs (cf. Section~\ref{sec:FCCHPmodel}). 
However, it is shown that for large fleets the MIP approach is impractical. Therefore a local search method was 
developed for the FPP, based on a dynamic programming formulation of the SHPP.  

For additional flexibility to freely combine components in a modular MIP model of an energy system and to add 
sub-models of further energy system components we categorized the energy system components and used some 
conventions in our modular and extendable MIP modeling approach (cf. Section~\ref{sec:optModel}). This approach was 
motivated by~\cite{guModelingPlanningOptimal2014}.

In~\cite{guOptimalConfigurationAnalysis2015} the optimal configuration and operation of  combined cooling, heating, 
and power (CCHP) microgrids are considered. Similar to our approach the uncertainty of cooling, heating, and power load 
is predicted, however, be aware the energy consumption can still deviate from the predicted values. The components of 
the microgrid considered in~\cite{guOptimalConfigurationAnalysis2015} and by us overlap in photovoltaic (PV) systems, 
(gas) boilers, thermal storage tanks (TSTs), absorption chillers, electric chillers, as well as in cooling, heating, 
and power loads. In~\cite{guOptimalConfigurationAnalysis2015} gas turbines and electric chillers are part of the 
microgrid while we take fcCHPs, batteries and heat or cold pumps into account, too. However, for optimization 
in~\cite{guOptimalConfigurationAnalysis2015} a nonlinear programming model is proposed, which aims to minimize the 
total costs of the CCHP system.  

\section{Developing Extendible, Modular Optimization Models}\label{sec:optModel}

In order to realize the optimizing component of an energy management system, we carried out an extendible, modular 
modeling approach of building energy systems. Therefore combinable MIP sub-models of the energetic behavior of plant 
components (cf. Section~\ref{sec:RW}) are developed. The optimization component generates corresponding mathematical 
optimization problems from problem-specific descriptions of building energy systems. Their solutions result in 
timetables resp. operation schedules for the components of the respective building energy system where the supply, use 
and conversion of the various forms of energy is determined. Real or fictitious costs (e.g., for \COO~emissions) 
can be minimized by this component. 

A modular modeling approach is chosen where each component is characterized by whether it is an electrical, warming, 
cooling or financial source, sink or reservoir.\footnote{In this context ``modular'' means that the MIP sub-models can 
be combined in accordance to any building energy system specification -- not in the sense of 
\cite{jarvisaloModuleBasedFrameworkMultilanguage2009}.} For example, energetic sources provide power, energetic sinks 
draw it. Energetic reservoirs  have an energy level and can both draw and deliver power within minimum and maximum 
levels. With this knowledge, it is then determined for each time unit in the scheduling horizon that the sum of power 
supplies and demands must be balanced, i.e., zero, for electricity, heat and cold. The sum of the yields and costs of 
the financial sources and sinks form the objective function for the optimization. With these conditions, the specific 
sub-models of the energy system components are combined to form an overall model. The sub-models, which are to be 
defined for each component under consideration of these conventions, describe the plant-specific energetic behavior as 
well as the associated financial effects. This approach was motivated by~\cite{guModelingPlanningOptimal2014}. Within 
this approach energy storage devices such as batteries are reservoirs. Energy converters such as heat pumps are both 
electrical sinks and thermal sources (either cold or heat, depending on the operating mode). In order to join the 
sub-models of the energy system components into an overall model of the energy system, a naming convention is used for 
decision variables that define the consumption or production of the respective energy type in a discrete unit of time. 
For example, any electrical source~$p$ has a variable ``$\f{electricOutputPower}_p(t)$'' specifying the electric power 
supply (output) during time unit~$t$. The integration of all electrical sinks and sources in an energy system model then 
takes place depending on the relevant system components with the help of the equation
\begin{eqnarray*}
  \forall t: \sum_{p \in \fs{ElectricalSource}} \f{electricOutputPower}_p(t) 
	  & = & \sum_{c \in \fs{ElectricalSink}} \f{electricInputPower}_c(t) \enspace,
\end{eqnarray*}
which states that the sum of electrical energy consumption and production must be balanced at all times. Similar 
equations are used to integrate system components via other types of energy. Since components are included in 
several equations, a quasi-automatic cross-sector coupling occurs across the considered energy types.

Since costs but also yields were used to optimize energy system operations, we have supplemented the chosen approach 
with financial sources and sinks. Their outputs (yields) and inputs (costs) are added together over the scheduling 
horizon 
\begin{eqnarray*}
  \sum_{t \in \fs{horizon}}\Big(\sum_{p \in \fs{FinancialSource}} \f{financialOutput}_p(t) 
	  & + & \sum_{c \in \fs{FinancialSink}} \f{financialInput}_c(t)\Big) \enspace,
\end{eqnarray*}
such that the optimization of an energy system is done either by minimizing the total costs or by maximizing the total 
yield, depending on whether the costs are represented by positive values and the yields by negative values or vice 
versa.  

Taking these characterizations and conventions into account, a set of extendible and connectable MIP sub-models were 
created for the following energy system components:
\begin{itemize}
	\item User behavior with time-variable electricity, heat/cooling and hot water requirements, 
	\item Mains connections with power limitations, time-variable electricity prices and refunds,
	\item Mechanical block-type CHPs with switchable peak load boilers with efficiency factors,
	\item Heat/cooling pumps with variable (outdoor temperature-dependent) Coefficients of Performance (COP),
	\item Heat/cold storages with charging losses and efficiency factors,
	\item Battery storages with charging losses and efficiency factors,
	\item Absorption chillers with efficiency factors,
	\item Heating rods and burners with efficiency factors,
	\item Photovoltaic (PV) systems with predicted power supply,
	\item fcCHPs with their special characteristics.
\end{itemize}
In this context efficiency factors ($\in [0,1]$) are reflecting energy conversion losses.
Modeling approaches from~\cite{bozchaluiOptimalOperationCommercial2012} and useful suggestions for MIP modeling 
coming from~\cite{beasleyNotesSeparableProgramming} are adopted. In addition to the characteristic energetic behavior, 
cost factors such as (variable) primary energy costs or costs for emissions as well as wear and tear 
costs during start-up and shut-down of plants, i.e., operating and maintenance costs, were also taken into account. 
The most challenging part was the modeling of fcCHPs with their special characteristics. With the help of a fcCHP 
manufacturer, we created a mathematical model to describe the energetic relationships in fcCHPs. This will be presented 
in detail in the next section. 

\subsection{A MIP Model for Fuel Cell Combined Heat and Power Plants}\label{sec:FCCHPmodel} 

\begin{figure}
  \begin{center}
	  \vspace{-4mm}
    \includegraphics[width=1.0\textwidth]{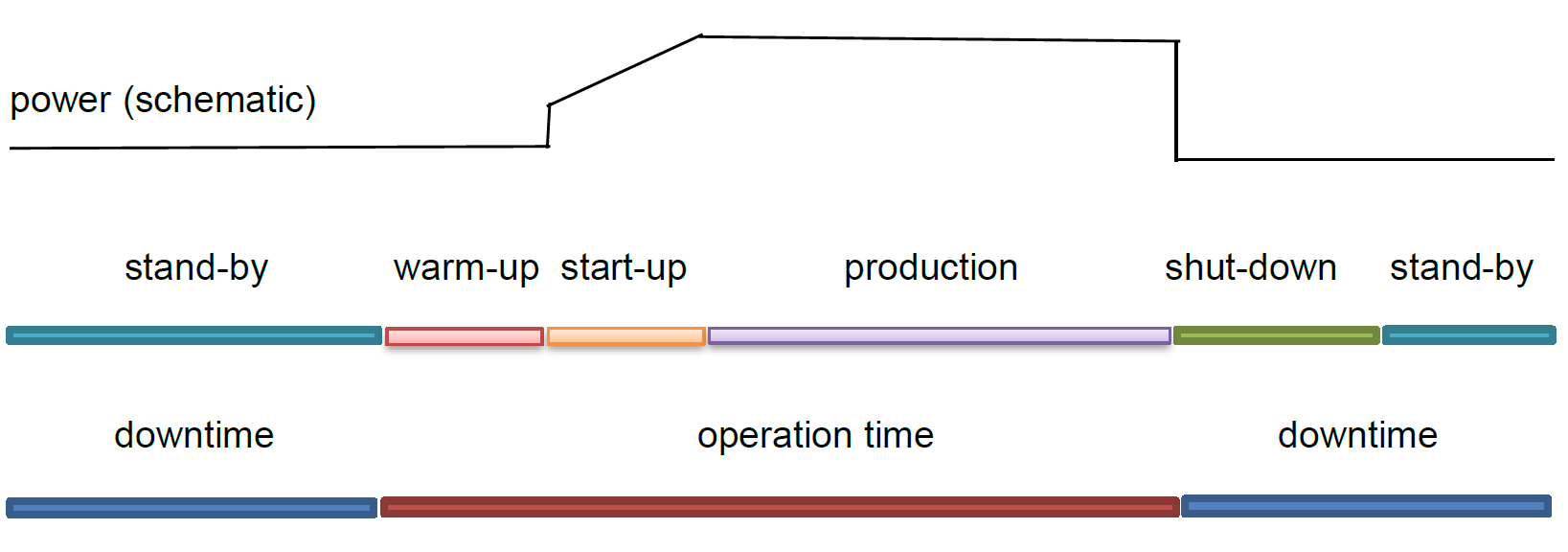}
    \caption{Schematic power profile and according state phases of a fcCHP.}
    \label{fig:fcCHPschema}
		\vspace{-8mm}
  \end{center}
\end{figure}

FcCHPs have characteristic physical parameters (constant values) and characteristic curves for broad electrical energy, 
thermal energy and primary energy supply on the basis of monitoring data from practical tests. 
For fcCHPs their processing phases such as cold start, warm start etc. as well as their power modulation opportunities 
are typical. The individual phases within downtime and operating time are shown in Figure~\ref{fig:fcCHPschema}. 
For example, the provision of thermal and electrical energy is delayed by a warm-up phase with a duration depending 
on the length of the immediately preceding downtime. Furthermore, typical consumption data for primary energy 
(e.g., natural gas) and electric energy were given by the manufacturer on the basis of measurements  during the 
individual phases.
On the basis of characteristic parameters and curves of fcCHPs as well as explanations of the corresponding energetic 
correlations, we developed a general mathematical model which formally describes the relationship between primary 
energy demand as well as thermal and electrical energy supply. Therefore, the model distinguishes between different 
phases and explains the temporal dependencies between these phases. Within our model of a fcCHP we used the following 
physical parameters (constant values) of such power plants:
\begin{itemize}
  \item A \emph{thermal efficiency factor} $0 < \eta_{\fs{th}} < 1$ and an \emph{electric efficiency factor} $0 < 
	  \eta_{\fs{el}} < 1$ (with respect to the primary energy source) such that $\eta_{\fs{th}} + \eta_{\fs{el}} < 1$.
	\item A \emph {maximal power output (thermal/electric) within the production phase} $P_{\fs{th}_{\max}}$ resp. 
	  $P_{\fs{el}_{\max}}$ where $\eta_{\fs{el}} \cdot P_{\fs{th}_{\max}} = \eta_{\fs{th}} \cdot P_{\fs{el}_{\max}}$ 
		because in general it applies that $P_{\fs{th}} / \eta_{\fs{th}} = P_{\fs{prim}} =  P_{\fs{el}} / \eta_{\fs{el}}$ 
		where $P_{\fs{prim}}$ is the power of the primary energy carrier.
	\item A \emph {minimal power output (thermal/electric) within the production phase} $P_{\fs{th}_{\min}}$ resp. 
	  $P_{\fs{el}_{\min}}$ where $\eta_{\fs{el}} \cdot P_{\fs{th}_{\min}} = \eta_{\fs{th}} \cdot P_{\fs{el}_{\min}}$.
	\item A \emph{minimal and maximal operation time} $D_{\fs{on}_{\min}}$ resp. $D_{\fs{on}_{\max}}$ as well as
	  a \emph{minimal off-time} $D_{\fs{off}_{\min}}$.
	\item a \emph{bounded, monotonically increasing function~$f: \mathbb{N}^+ \to \mathbb{N}^+$ to determine the warm-up 
	  time}~$d_{\fs{warmUp}}$ depending on its recent \emph{off-time}~$d_{\fs{off}}$, i.e., $d_{\fs{warmUp}} = 
	  f(d_{\fs{off}})$. 	
	\item A \emph{constant electric power input during stand-by phases}: $P_{\fs{el}_{\ft{standBy}}}$.
	\item Some \emph{constant primary and electric power inputs} during the warm-up phase: $P_{\fs{pr}_{\ft{warmUp}}}$ 
	  and $P_{\fs{el}_{\ft{warmUp}}}$. There, a cold-start requires additional input power: 
		$P_{\fs{pr}_{\ft{coldStart}}}$ and $P_{\fs{el}_{\ft{coldStart}}}$.
	\item An \emph{additional constant electric power input during shut-down}: $P_{\fs{el}_{\ft{addShutDown}}}$ during 
	  the short shut-down interval~$D_{\fs{down}}$. This means that electric power input from the production phase to 
		the stand-by phase has the power peak $P_{\fs{el}_{\ft{standBy}}} + P_{\fs{el}_{\ft{addShutDown}}}$.
  \item A \emph{constant thermal and electric output power ``peak''} at the beginning of the start-up phase from zero 
	  to~$P_{\fs{th}_{\ft{init}}}$ resp. to~$P_{\fs{el}_{\ft{init}}}$ within a (short) constant time interval 
		$D_{\fs{init}}$ where $\eta_{\fs{el}} \cdot P_{\fs{th}_{\ft{init}}} = \eta_{\fs{th}} \cdot P_{\fs{el}_{\ft{init}}}$
		applies.
	\item A \emph{constant power enhancement within a constant start-up phase to a final power value}: It is assumed that
	  the total duration of the start-up phase~$D_{\fs{startUp}}$ as well as the final power 
		value~$P_{\fs{th}_{\ft{startUp}}}$  are given such that $P_{\fs{th}_{\ft{init}}} \le P_{\fs{th}_{\ft{min}}} < 
		P_{\fs{th}_{\ft{startUP}}} \le P_{\fs{th}_{\ft{max}}}$ applies. Thus, the constant power enhancement is
		$(P_{\fs{th}_{\ft{startUp}}} - P_{\fs{th}_{\ft{init}}}) / (D_{\fs{startUp}} - D_{\fs{init}})$. Consequently, the
		electric power enhancement results from $P_{\fs{el}_{\ft{startUp}}} = \eta_{\fs{el}} / \eta_{\fs{th}} \cdot
		P_{\fs{th}_{\ft{startUp}}}$.
	\item A \emph{maximal gradient for power modulation in the production phase}: $\Delta P_{\fs{th}_{\ft{prod}}} / 1h$ 
	  resp. $\Delta P_{\fs{el}_{\ft{prod}}} = \eta_{\fs{el}} / \eta_{\fs{th}} \cdot \Delta P_{\fs{th}_{\ft{prod}}}$.
\end{itemize}
The chosen MIP model of fcCHPs uses discrete time units. Therefore, the considered scheduling horizon $[0, T]$ is 
divided into $N$~equidistant time intervals of equal duration -- typically of 15~min\footnote{A one-day scheduling 
horizon is subdivided into 96 time units.}, however, other time unit durations are possible, too. It is assumed that a 
fcCHP is either in operation or down within one time unit. Discretization of time is a common approach in mathematical 
modeling of dynamic processes. Further, it is compatible with the time units used in short-term energy markets. This 
means that a scheduling horizon~$[0,T]$ is divided into $N$ time units / intervals $[t_{i-1},t_i) (i=1,\ldots,n)$ 
of the same duration, namely $T/N$.
 
The minimum and maximum operating durations (in time units) are therefore $\f{On}_{\min} = \lceil D_{\fs{on}_{\min}} 
\cdot N/T \rceil$ and $\f{On}_{\max} = \lceil D_{\fs{on}_{\max}} \cdot N/T \rceil$ and the minimum downtime is 
$\f{Off}_{\min} = \lceil D_{\fs{off}_{\min}} \cdot N/T \rceil$ time units.

The decision whether a fcCHP is switched on or off (operation time vs. downtime) is always made for a complete 
time unit~$i$ (i.e., for a time interval $[t_{i-1}, t_i)$). For this purpose, Boolean decision variables $x_0,...,x_N$ 
are introduced and $x_i = 1$ applies if the fcCHP is \emph{on} at time unit~$i$ and $x_i = 0$ if it is \emph{off} at 
time unit~$i$, where $x_0$ indicates the on/off state at the beginning of the scheduling horizon which is known in 
advance. Furthermore, for $i=2-\f{On}_{\min}, \ldots, N$ the \emph{start variables} $\f{start}_i$ are Boolean decision 
variables which determine whether the fcCHP starts in time unit~$i$ (start of the operating phase) or not, i.e., 
$x_i = 1$ and $x_{i-1} = 0$ applies or not. There, for $j=2-\f{On}_{\min}, \ldots, 0$ $\f{start}_i$ indicates any 
potentially interesting start event in the past which is known in advance. Similarly, the \emph{stop variables} 
$\f{stop}_i$ are Boolean decision variables which determine whether the fcCHP is switched off in time unit~$i$ (begin 
of the down phase) or not, i.e., $x_i = 0$ and $x_{i-1} = 1$ applies or not.

In order to ensure that the start and stop variables are compatible with the on/off variables, the following 
conditions must be met (cf.~\cite{bosmanPlanningProductionFleet2012}). There, the status of the fcCHP immediately 
before the start of the scheduling horizon, namely~$x_0$, is relevant:
\begin{eqnarray*}
  \begin{array}{rclcrcl}
    \f{start}_i & \ge & x_i - x_{i-1} 
		  & \qquad & \f{stop}_i & \ge & x_{i-1} -x_i \\ 	
    \f{start}_i & \le & x_i 
		  & \qquad & \f{stop}_i & \le & x_{i-1} \\ 	
	  \f{start}_i & \le & 1- x_{i-1} 
		  & \qquad & \f{stop}_i & \le & 1 - x_i
	\end{array}	
	& \qquad & \mbox{for $i = 1, \ldots, N$.}
\end{eqnarray*}
In order to further ensure that neither the minimum operating times nor downtimes are undercut, the following 
conditions must also be fulfilled:
\begin{eqnarray*}
 \begin{array}{rclcrcl} 
   \displaystyle x_i & \displaystyle\ge & \displaystyle \sum_{k=i-\fs{On}_{\min}+1}^{i-1} \f{start}_k 
	 & ~\displaystyle \land~ & 
   \displaystyle x_i & \displaystyle\le & \displaystyle 1 - \sum_{k=i-\fs{Off}_{\min}+1}^{i-1} \f{stop}_k 
	\end{array}	
	& \qquad & \mbox{for $i = 1, \ldots, N$.}
\end{eqnarray*}
\begin{example}
Let a fcCHP with a minimal operation time~$\f{On}_{\min}=5$ time units be given. Further let~$\f{start}_{-3} = 1$ and
$\f{start}_{-2} = \f{start}_{-1} = \f{start}_{0} = 0$. Then for any admissible schedule $x_1 = 1$ must apply, i.e.,
the fcCHP must be operative (``on'') at time unit~1, otherwise its minimal operation time is undercut. 
\end{example}

In order to limit the operating time, further auxiliary integer variables $l_1, \ldots,l_N$ are required, such that 
the difference $l_i - l_{i-1}$ corresponds to the duration (in time units) from the last stop or start when starting or 
stopping at time unit~$i$ assuming that a stop follows a start and vice versa. For this purpose let $l_0 \le 0$ be the 
non-positive time unit at the \emph{last} start or stop just before the beginning of the considered scheduling horizon.
For any time unit~$i \in \{1, \ldots, N\}$ the last start/stop time unit is kept if the on/off status of the fcCHP
doesn't change: If $x_i = x_{i-1}$ applies, then let $l_i = l_{i-1}$. Otherwise, if there is change of the status the
last start/stop time unit is updated: If $ x_i \ne x_{i-1}$ applies, then let $l_i = i$. Combining both cases results 
in:
\begin{eqnarray}  
	l_i & = & (1-|x_i - x_{i-1}|) \cdot l_{i-1} + |x_i - x_{i-1}| \cdot i \qquad\mbox{for $i = 1, \ldots, N$.}
	\label{eqAbs}
\end{eqnarray}

\begin{example}
Let a fcCHP be given which runs from time unit~-3 (already running at the beginning of the scheduling horizon) 
to time unit~13. Consequently~$l_0 = -3$, $x_{0} = x_1 = \cdots = x_{13} = 1$ and $x_{14} = 0$ apply. Thus, $l_1 =
\cdots = l_{13} = -3$ but $l_{14} = 14$ apply due to the fact that $x_{13} = 1$ and $x_{14} = 0$. Then the difference
$l_{14} - l_{13} = 14 - (-3) = 17$ defines the recent operation time of the fcCHP in time units.
\end{example}

In general, Equation~(\ref{eqAbs}) cannot be processed directly by a MIP-Solver, because it contains products of
Boolean terms and decision variables. Therefore any such product $\alpha \cdot U$ with $\alpha \in \{0, 1\}$ and 
$U \in [u_{\min}, u_{\max}]$ has to be replaced by a new auxiliary decision variable $V \in [\min(0, u_{\min}), 
u_{\max}]$ and the additional linear inequalities
\begin{eqnarray*}
  u_{\min} \cdot \alpha \le V & \land & V \le u_{\max} \cdot \alpha \quad \land  \\
	U - u_{\max} \cdot (1 - \alpha) \le V  & \land & V \le U - u_{\min} \cdot (1 - \alpha) \enspace.
\end{eqnarray*}
The replacement is correct: 
On the one hand it follows from $\alpha = 0$ that $U - u_{\max} \le 0 \le V \le 0 \le U - u_{\min}$  applies and therefore $V = 0$. On the other hand if follows from $\alpha = 1$ that $u_{\min} \le U \le V \le U \le u_{\max}$ applies and therefore $V = U$.  In summary, $V = \alpha \cdot U$ applies.   

Furthermore, Equation~(\ref{eqAbs}) contains the absolute amount of a difference. However, any equation $X = |B-A|$ 
can be modeled by means of a new auxiliary Boolean variable~$\beta \in \{0, 1\}$ and some additional linear constraints
\begin{eqnarray*}
	X & \ge & 0 \quad\land \\
	X &  =  & \beta \cdot (B - A) + (1 - \beta) \cdot (A - B) \enspace.
\end{eqnarray*}
Consequently, either $X = A - B$ or $X = B - A$ applies depending on the value of~$\beta$. Since $X$ must not be negative, $X = |B-A| = |A-B|$ applies.

In order to ensure that the maximum operating time is not exceeded, the following must therefore apply:\footnote{%
Here and in the following there are products of Boolean terms and decision variables, too.}
\begin{eqnarray*}
  \f{stop}_i \cdot (l_i - l_{i-1}) & \le & \f{On}_{\max} \qquad\mbox{for $i = 1, \ldots, N$.}
\end{eqnarray*}
These auxiliary variables are also useful to determine the duration of downtimes, which will be $\f{start}_i \cdot (l_i 
- l_{i-1})$ and thus the duration of warm-up times, which will be $f(\f{start}_i \cdot (l_i - l_{i-1}))$.\footnote{%
Remember that the function~$f$ maps downtimes to warm-up times, see above.}

If the downtime is greater than a specified value~$L > 0$, this is referred to as a \emph{cold start}. Auxiliary 
Boolean variables $k_1, \ldots, k_N$ are given, such that the value of~$k_i$ in the warm-up phase indicates 
whether this occurred after a cold start, i.e., $k_i=1$ is implied:
\begin{eqnarray*}
  \f{start}_i \cdot (i - l_{i-1}) - L \le  M \cdot k_i & \land &  k_{i-1} \cdot (i - l_i) \le  M \cdot k_i 
\end{eqnarray*}
for $i = 1, \ldots, N$, a sufficiently large value~$M$ and a corresponding value $k_0$, e.g., known from a previous 
scheduling horizon. If the modeled fcCHP is starting at time unit~$i$ then $\f{start}_i = 1$ and $l_i = i$ 
apply. If this start event is a cold start, i.e., if $(i - l_{i-1}) > L$ applies, then $\f{start}_i \cdot 
(l_i - l_{i-1}) - L > 0$ applies, too. If follows that $k_i = 1$ applies, otherwise there is a violation. 
If $k_{i-1} = 1$ indicates that a downtime will require a cold-start and there is not any start event at 
time~$t_i$ then $k_i = 1$ is implied further indicating a cold-start, because $(i - l_i) > 0$ applies, 
otherwise there is violation.

The determination of the warm-up times requires additional auxiliary integer variables $w_1, \ldots,w_N$, such that 
the value of~$w_i$ corresponds to the last \emph{warm-up} time. To do this, let $w_0$ be the warm-up time from the 
previous scheduling horizon. If $\f{start}_i = 1$ applies, let $w_i = f(l_i - l_{i-1})$. Otherwise, 
if $\f{start}_i = 0$ applies, let $w_i = w_{i-1}$. Combining both cases results in
\begin{eqnarray*}
  w_i & = & \f{start}_i \cdot f(i - l_{i-1}) + (1 - \f{start}_i) \cdot w_{i-1} \qquad\mbox{for $i = 1, \ldots, N$.}
\end{eqnarray*}
Due to the fact that the argument of the function~$f$ is variable, i.e., not known in advance, the computation 
of~$f(x)$ for a variable $x \in \{1, \ldots, N\}$ (assuming that the maximum downtime is shorter than the scheduling 
horizon) requires additional auxiliary Boolean variables $\lambda_1, \ldots, \lambda_N$. Then the condition
\begin{eqnarray*}
  \forall i \in \{1, \ldots, N\} : \lambda_i \cdot (x - i) = 0  & \land & \sum_{i=1}^N \lambda_i = 1
\end{eqnarray*}
ensures that $x = i \Leftrightarrow \lambda_i = 1$ applies for $i = 1, \ldots, N$. Consequently, it applies 
\begin{eqnarray*}
  f(x) = \sum_{i=1}^N \lambda_i \cdot F_i \qquad\mbox{for any $x \in \{1, \ldots, N\}$,}
\end{eqnarray*}
where the supporting values $F_1 = f(1), \ldots, F_n = f(N)$ are technical parameters of the fcCHP which are known in 
advance.\footnote{$F_1 \le \cdots \le F_n$ applies due to the fact that~$f$ is monotonically increasing, see above.} 

Furthermore, Boolean decision variables $y_1,...,y_N$ are introduced such that $y_i = 1$ applies if and only if 
the modeled fcCHP warms up in time unit~$i$. In particular it applies that $x_i \ge y_i$.
Additionally, $\f{stopWarmUp}_i$ are Boolean decision variables that determine whether the fcCHP has completed 
the end of the warm-up phase in time unit~$i$ (i.e., the start of the production phase) or not, i.e., $y_{i-1} = 1$ 
and $y_{i} = 0$ apply. In order to ensure that these ``stop of warm-up'' variables are compatible with the 
``warm-up'' variables, the following conditions must be met, whereby the ``warm-up'' state of the fcCHP 
immediately before the start of the scheduling horizon -- determined by~$y_0$ -- is relevant:
\begin{eqnarray*}
  \begin{array}{rclcrcl}
	  \f{start}_i & \ge & y_i - y_{i-1}
	    & \qquad & \f{stopWarmUp}_i & \ge & y_{i-1} - y_i \\
	  \f{start}_i & \le & y_i
		  & \qquad & \f{stopWarmUp}_i & \le & y_{i-1} \\ 		
	  \f{start}_i & \le & 1 - y_{i-1}
	    & \qquad & \f{stopWarmUp}_i & \le & 1 - y_i 	
	\end{array}	
	& \qquad & \mbox{for $i = 1, \ldots, N$.}
\end{eqnarray*}
A minimal duration of the warm-up phase has to be guaranteed. Therefore for each time unit~$i = 1, \ldots, N$ and 
for each possible warm-up duration~$j = F_1, \ldots ,F_n$ an auxiliary Boolean variable~$\sigma_{i,j}$ is defined
such that $\sigma_{i,j} = 1$ if and only if a start occured no longer than $j$~time units before time unit~$i$:
\begin{eqnarray*}
  \sigma_{i,j} \cdot M \ge \sum_{k=i-j+1}^{i-1} \f{start}_k & \land & \sigma_{i,j} \le \sum_{k=i-j+1}^{i-1} \f{start}_k 
\end{eqnarray*}
Then the minimal warm-up time is satisfied, if 
\begin{eqnarray*}
  (F_n - F_1 + 1) \cdot y_i & \ge & (w_i - j + 1) \cdot \sigma_{i,j}
\end{eqnarray*}
applies for $i = 1, \ldots, N$ and $j = F_1, \ldots ,F_n$. This means that if $w_i - j + 1$ is positive and the start 
is no longer than $w_i$ time units ago, i.e., $\sigma_{i,j} = 1$, then the fcCHP is in the warm-up phase, i.e., 
$y_i = 1$ must apply.\footnote{N.B.: $(F_n - F_1 + 1) \ge (w_i - j + 1)$ always applies, see above.}  

For an upper boundary of the warm-up time, additional auxiliary integer variables are necessary. Let $r_1, \ldots,r_N$ 
be given such that~$r_i$ represents the index of the last (i.e., most recent) start. Therefore let $r_0 \le 0$ be 
the time unit of the latest start before the scheduling horizon. Now if $\f{start}_i = 1$ then $r_i = i$ will apply, 
otherwise $r_i = r_{i-1}$:
\begin{eqnarray}
  r_i & = & \f{start}_i \cdot i + (1 - \f{start}_i) \cdot r_{i-1}
\end{eqnarray}
Then the maximal warm-up time is satisfied, if 
\begin{eqnarray}
  y_i \cdot i - r_i \le w_i \enspace.
\end{eqnarray}
This means that if the fcCHP warms up at time unit~$i$, then the latest start is no longer than the warm-up
time ago.

At the end of the warm-up phase, the start-up phase begins, the duration of which is known in advance from the 
fcCHP characteristics. The same applies to the thermal and electrical power available in the start-up phase. 
In detail, there are $\f{LowerInit} = \lfloor D_{\fs{init}} \cdot N/T \rfloor$ time units with power jump, in general
one time unit at $\f{UpperInit} = \lceil D_{\fs{init}} \cdot N/T \rceil$ with parts of the power jump and gradual 
starting (if $\f{\f{\f{\f{LowerInit}}}} < \f{UpperInit}$ and then $\f{StartUp} - \f{UpperInit}$  time units 
in which the power increases constantly up to a given target value with $\f{StartUp} = \lceil D_{\fs{startUp}} \cdot 
N/T \rceil$. The end of the start-up phase is thus after further $\f{StartUp}$ time units reached. This means that 
discrete power levels can be determined for primary energy and electricity consumption as well as for thermal and 
electric output power (abstract $P_{x\fs{Up}}$). One type is sufficient, the others behave proportionally according 
to their efficiency factors:
\[
  (P_{x\fs{Up}_1}, \ldots, P_{x\fs{Up}_{\ft{StartUp}}}) \enspace.
\]
Analogously to these power steps and due to the discretization there result electric power steps from 
$P_{\fs{el}_{\ft{addShutDown}}}$ during the shut-down phase (mostly one time unit because the duration~$D_{\fs{Down}}$
of the shut-down phase is in general short):
\[
  (P_{\fs{el}_{\ft{Down}_1}}, \ldots, P_{\fs{el}_{\ft{Down}_{\ft{ShutDown}}}}) \enspace,
\]
where $\f{ShutDown} = \lceil D_{\fs{Down}} \cdot N/T \rceil$ is the duration of the shut-down phase in time units.

The time units of the ``jump'' phases are characterized by Boolean decision variables $s_1, \ldots ,s_N$ where 
$s_i = 1$, if the fcCHP makes a performance jump in time unit~$i$ and $s_i = 0$, if it is not the case
 in this time unit~$i$:
\[
  s_i = \sum_{j=1}^{\fs{LowerInit}} \f{stopWarmUp}_{i-j+1} \enspace.
\]
Boolean decision variables $z_1, \ldots, z_N$ are introduced for the following production phase. $z_i = 1$ will apply, 
if the fcCHP is productive in time unit~$i$, i.e., delivering thermal and electrical power and $z_i = 0$ will apply,
if it is not in the production phase in time unit~$i$, i.e., in particular, it applies that $x_i \ge z_i$.
In order to ensure that these ``productive'' variables are compatible with the corresponding start/stop variables, the 
following conditions must be met, whereby the status of the fcCHP directly before the start of the scheduling 
horizon -- determined by~$z_0$ -- is relevant:
\begin{eqnarray*}
  \begin{array}{rclcrcl}
	  \f{stopWarmUp}_{i-\fs{StartUp}} & \ge & z_i - z_{i-1}  
	    & \qquad & \f{stop}_i & \ge & z_{i-1} - z_i \\ 	
	  \f{stopWarmUp}_{i-\fs{StartUp}} & \le & z_i 
		  & \qquad & \f{stop}_i & \le & z_{i-1} \\ 		
	  \f{stopWarmUp}_{i-\fs{StartUp}} & \le & 1 - z_{i-1}
	    & \qquad & \f{stop}_i & \le & 1 - z_i 	
	\end{array}	
	& & \mbox{for $i=1,\ldots,N$.}
\end{eqnarray*}

Summarizing, the \emph{thermal power supply (output)} of a fcCHP at time unit~$i$ is characterized by the 
following equation:   
\begin{eqnarray*}
  \f{thermalOutputPower}_{\fs{fcCHP}_i} & = & \sum_{j=1}^{\fs{StartUp}} 
	\f{stopWarmUp}_{i-j+1} \cdot P_{\fs{th}_{\ft{Up}_j}} + z_i \cdot u_{\fs{th}_i}
\end{eqnarray*}
where the values of the variable~$u_{\fs{th}_i}$ must lie within a specified performance band in the production phase, 
i.e., $P_{\sf{th}_{\min}} \le u_{\fs{th}_i} \le P_{\sf{th}_{\max}}$ and the gradient of the value change is limited:%
\footnote{How to model the absolute amount of a difference has already been explained.}
\begin{eqnarray*}
  |u_{\fs{th}_i} - u_{\fs{th}_{i-1}}| & \le & \Delta P_{\sf{th}_{\ft{prod}}}/1h \cdot T/N \enspace.
\end{eqnarray*}

The electrical power supply (output) results directly from the thermal power supply:
\begin{eqnarray*}
	\f{electricOutputPower}_{\fs{fcCHP}_i} 
	= \frac{\eta_{\fs{el}}}{\eta_{\fs{th}}} \cdot \f{thermalOutputPower}_{\fs{fcCHP}_i}
\end{eqnarray*}
The \emph{electrical power demand (input)} of a fcCHP 
depends on whether there is a cold-start or a warm-start:
\begin{eqnarray*}
  \f{electricInputPower}_{\fs{fcCHP}_i} 
	& = & y_i \cdot P_{\sf{el}_{\ft{WarmUp}}} + (y_i \land k_i)  \cdot P_{\fs{el}_{\ft{ColdStart}}} \\
	&   & + \sum_{j=1}^{\fs{ShutDown}} \f{stop}_{i-j+1} \cdot P_{\fs{el}_{\ft{Down}_j}} 
	+ (1 - x_i) \cdot P_{\fs{el}_{\ft{standBy}}} \enspace.
\end{eqnarray*}
There, the conjunction of two Boolean variables~$(y_i \land k_i)$ will be represented by an auxiliary Boolean 
variable~$\gamma_i$ satisfying $\gamma_i \ge y_i + k_i - 1 \land \gamma_i \le y_i \land \gamma_i \le k_i$
for $i=1, \ldots, N$.

\begin{figure}
  \begin{center}
		\vspace{-4mm}
    \includegraphics[width=1.0\textwidth]{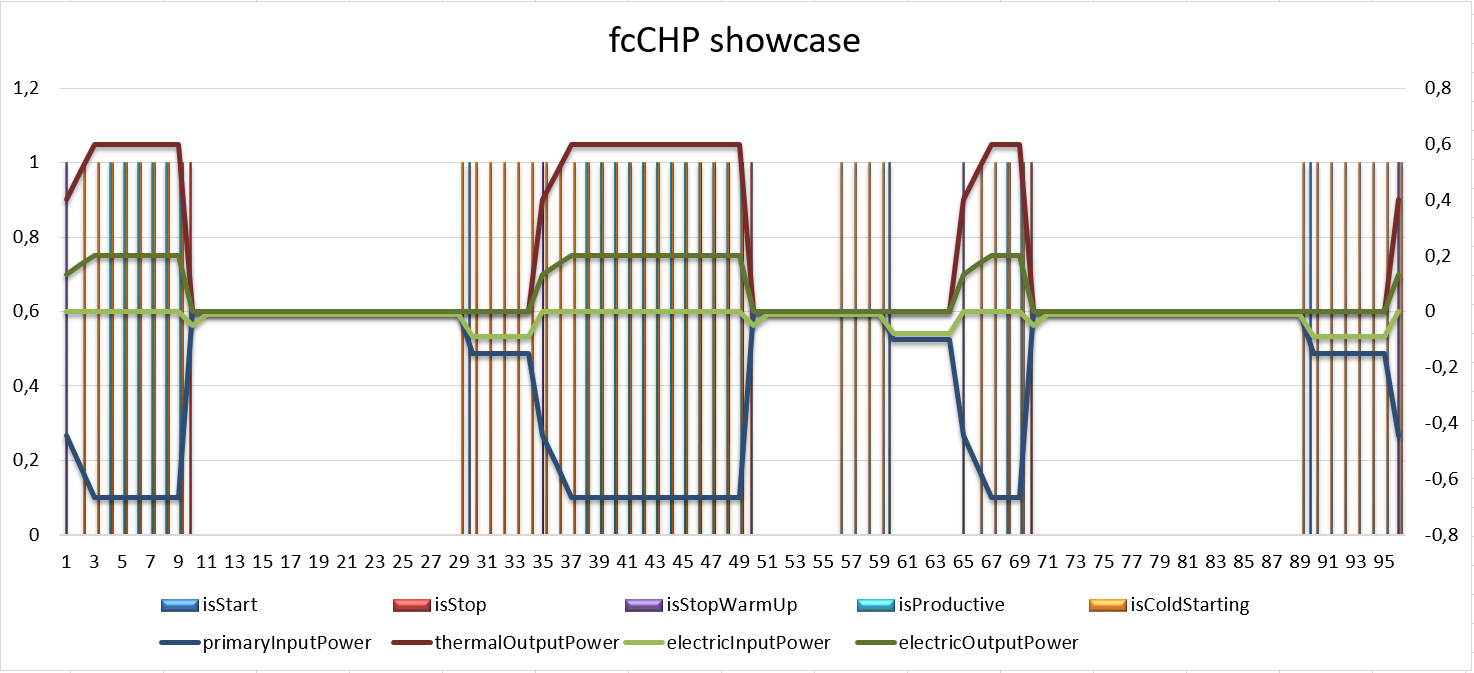}
    \caption{Typical operation of a fcCHP.}
    \label{fig:fcCHPshowcase}
		\vspace{-8mm}
  \end{center}
\end{figure}

Analogously, the \emph{primary power demand (input)} of a fcCHP 
over the production phases depends also on whether there is a cold-start or a warm-start:
\begin{eqnarray*}
  \f{primaryInputPower}_{\fs{fcCHP}_i} 
	& = & y_i \cdot P_{\sf{pr}_{\ft{WarmUp}}} + \gamma_i \cdot P_{\fs{pr}_{\ft{ColdStart}}} \\
	&   & + \sum_{j=1}^{\fs{StartUp}} \f{stopWarmUp}_{i-j+1} \cdot P_{\fs{pr}_{\ft{WarmUp}_j}} 
	+ z_i \cdot \frac{u_{\fs{th}_i}}{\eta_{\fs{th}}} \enspace.
\end{eqnarray*}
In Figure~\ref{fig:fcCHPshowcase}, typical thermal and electric power profiles of a fcCHP are shown according to the 
presented MIP model. These profiles are matching the profiles measured by the fcCHP manufacturer giving some evidence 
that the energetic behavior of fcCHP is modeled adequately.    
Assuming that the costs $K_{\fs{pr}_i}$ for primary energy at time unit~$i$, as well as the costs for switching on 
$K_{\fs{on}}$ and for switching off $K_{\fs{off}}$ a fcCHP, as well as the costs for the wear and tear per time unit 
during heating up $K_{\fs{warmUp}}$, during cold start $K{\fs{coldStart}}$ and during productive operation 
$K_{\fs{prod}}$ are known, then the costs for the operation of a fcCHP in time unit~$i$ result directly:
\begin{eqnarray*}
  \lefteqn{\f{financialInput}_{\fs{fcCHP}_i}} \\ 
	& = & K_{\fs{pr}_i} \cdot \f{primaryInputPower}_{\fs{fcCHP}_i} \cdot \frac{T}{N} + K_{\fs{on}} \cdot \f{start}_i \\
	&   & + K_{\fs{off}} \cdot \f{stop}_i + K_{\fs{warmUp}} \cdot y_i + K_{\fs{coldStart}} \cdot (y_i \land k_i) 
	      + K_{\fs{prod}} \cdot z_i  \enspace.
\end{eqnarray*}
This is only a simplified approach for the consideration of wear and tear costs of a fcCHP. For instance, aging 
effects are not taken into account. Although an aging approach can be converted into a linearly approximated 
model, initial run-time investigations result in very long computation times for cost optimization. However, the 
resulting operational plans hardly differ qualitatively from those with simplified models.

\section{Implementing a MIP-Based Optimizing Tool}\label{sec:MIP-Optimizer}

A software named ``MIP Optimizer'' is realized to transfer specifications of building energy systems into MIP 
models and then based on these models to determine cost-minimal operation schedules for the specified energy system 
components, so that predicted energy requirements for heating or cooling, domestic hot water and electricity over a
given scheduling horizon are covered. In detail, the MIP Optimizer generates a MIP optimization problem from a formal description of the energy system components, i.e., the \emph{configuration} of the energy system and from a formal description of the demand and the environmental and operational \emph{situation} over the scheduling period. For this purpose, both formal descriptions determining the general \emph{configuration} and the current \emph{situation} are to be specified in XML files (cf.~\cite{w3cExtensibleMarkupLanguage}) and must comply with a fixed XML schema 
(XSD)~\cite{w3cXMLSchema}. Energy demand profiles and other time series predicting the environmental situation 
(e.g., fluctuating primary energy prices or volatile PV~power) shall be provided in files in Hierarchical Data Format 
(HDF5)~\cite{thehdfgroupHDFHome,folkOverviewHDF5Technology2011}, referred to in the XML descriptions.
Examples of the XML configuration and situation files are presented in Section~\ref{sec:Evaluation}.

Then, the MIP Optimizer uses the $<$Coliop$|$Coin$>$ Mathematical Programming Language (CMPL) 
\cite{steglichCMPLColiopMathematical2018} to generate and solve the optimization problem, since various MIP problem 
solvers can be used, such as the freely  available Cbc~\cite{forrestCBCUserGuide} or the commercial 
CPLEX~\cite{ibmILOGCPLEXOptimization2019}. The operation schedules of the components of the building energy system are 
then extracted from the solution and stored in the form of time series in an HDF5 file, such that these data can
be further used by a building management system to control the energy components. 

The MIP Optimizer is implemented in Java and has a modular structure. Due to the modular modeling approach (cf. 
Section~\ref{sec:optModel}) and its object-oriented implementation, flexible extensions including further 
energy systems components are supported by design. 

\section{Evaluation on a Heat Pump Scenario}\label{sec:Evaluation}

In order to prove the usefulness of our MIP-based optimization approach, generated operational schedules 
for building energy systems are considered. For this purpose, we considered a residential building, i.e.,
a single family detached house, according to the EnEV standard 2014 with $172~\mbox{m}^2$ usable area (Berlin site), 
an air-to-water heat pump with hot water storage tank for heat supply. 

It was investigated, how an efficient operation can be planned with the help of the MIP Optimizer as cost- and 
energy-efficient as possible by predicting the Coefficients of Performance (COP) of the heat pump dependent 
on the outside temperature and the residential heat demand. The building energy system consists of a heat pump and 
a heat storage. The heat demand and COP of the heat pump changing with the outside temperature were determined and 
provided by our project partners at Berlin University of the Arts, Institute for Architecture and Urban Planning, 
Department of Building Physics and Building Technology. These partners used weather forecasts and Modelica to model 
and simulate the thermal behavior of the selected building.

In the subsequent operational scheduling, partial models of a heat pump with a constant electrical power consumption 
of 1.8~kW in operation (cf. the XML element {\tt <HeatPump>} below) and of a $2~\mbox{m}^3$ heat storage tank with 
a charging capacity of 20.82~kWh (cf. the XML element {\tt <HeatBuffer>} below) were combined at an ambient temperature 
of $20~\mbox{C}^{\circ}$ to form an overall model of the energy system. 
Electricity prices, the heat demand profiles determined by simulation and time-dependent COP as well as the system 
status data (e.g., state of charge of the heat storage tank) were added. The description of the building energy 
system was specified in XML as a \emph{configuration} with characteristic physical parameters, which were further
processed by the MIP Optimizer:



\lstset{
  keywords=[0]{BuildingConfiguration,BuildingSituation,Usage,Grid,HeatBuffer,HeatPump,ElectricPowerUsage,
	HotWaterPowerUsage,MinHeatingPowerUsage,MaxHeatingPowerUsage,MinCoolingPowerUsage,MaxCoolingPowerUsage,
	ElectricEnergyPrice,ElectricEnergyRefund,CoefficientOfPerformance},
  keywords=[1]{id,powerUnit,energyUnit,priceUnit,energyPriceUnit,electricPower,maxElectricPowerUse,maxHeatingPowerUse,
	maxCoolingPowerUse,maxFeedInPower,maxSupplyPower,minThermalEnergyLevel,maxThermalEnergyLevel,thermalLossPerHourFactor,
	maxThermalChargingPower,maxThermalDischargingPower,minOffTimeInHours,minRunTimeInHours,nbsOfTimeUnits,
	hoursPerTimeUnit,start,fileNameHDF5,maxInitialHeatingEnergy,maxInitialCoolingEnergy,fileName,dataSetPath,
	initialThermalEnergyLevel,isOnAtBegin,lastStartStopChangeInHours},
  keywordstyle=[0]\ttfamily\bfseries\color{Blue},
  keywordstyle=[1]\ttfamily\bfseries\color{RoyalBlue},
	basicstyle=\scriptsize\ttfamily\bfseries,
	stringstyle=\color{OliveGreen},
	tabsize=2,
}

\begin{lstlisting}
<BuildingConfiguration
    xmlns="http://www.fokus.fraunhofer.de/WaveSave"
    xmlns:xsi="http://www.w3.org/2001/XMLSchema-instance"
    xsi:schemaLocation="http://www.fokus.fraunhofer.de/WaveSave BuildingSystem.xsd"
    id="UDKHeatPumpScenario" powerUnit="kW" energyUnit="kWh" priceUnit="ct"
    energyPriceUnit="ct/kWh">
  <Usage id="generalUsage" maxElectricPowerUse="32.0" maxHeatingPowerUse="32.0" 
      maxCoolingPowerUse="0.0" powerUnit="kW"/>
  <Grid id="GridConnection" maxFeedInPower="0.0" maxSupplyPower="32.0" 
      powerUnit="kW"/>
  <HeatBuffer id="HotWaterBuffer" minThermalEnergyLevel="0" 
      maxThermalEnergyLevel="20.82" thermalLossPerHourFactor="0.000"
      maxThermalChargingPower="10.0" maxThermalDischargingPower="10.0"
      powerUnit="kW" energyUnit="kWh"/>
  <HeatPump id="HeatPump" electricPower="1.8" powerUnit="kW" 
			minOffTimeInHours="0.25" minRunTimeInHours="0.25"/>
</BuildingConfiguration>
\end{lstlisting}


Actual and predicted data were also transferred to the MIP Optimizer with the help of a \emph{situation} description, 
also in XML,  whereby the time series for heat demand, COP, electricity prices etc. were to be found in separate HDF5 
files to which references are made:
\begin{lstlisting}
<BuildingSituation
        xmlns="http://www.fokus.fraunhofer.de/WaveSave"
        id="UDKHeatPumpScenario" nbsOfTimeUnits="96" hoursPerTimeUnit="0.25" 
				start="2016-08-17T00:00:00" fileNameHDF5="UDKHeatPumpScenario.h5">
    <Usage id="generalUsage" maxInitialHeatingEnergy="0.0" 
			maxInitialCoolingEnergy="0.0" energyUnit="kWh">
        <ElectricPowerUsage fileName="UDK Heat Pump Scenario-2017-05.h5" 
					dataSetPath="/ENull" powerUnit="kW"/>
        <HotWaterPowerUsage fileName="UDK Heat Pump Scenario-2017-05.h5" 
					dataSetPath="/DHWNull" powerUnit="kW"/>
        <MinHeatingPowerUsage fileName="UDK Heat Pump Scenario-2017-05.h5" 
					dataSetPath="/MinHeating" powerUnit="W"/>
        <MaxHeatingPowerUsage fileName="UDK heat pump scenario-2017-05.h5" 
					dataSetPath="/MaxHeating" powerUnit="W"/>
        <MinCoolingPowerUsage fileName="UDK Heat Pump Scenario-2017-05.h5" 
					dataSetPath="/MinCoolingNull" powerUnit="kW"/>
        <MaxCoolingPowerUsage fileName="UDK Heat Pump Scenario-2017-05.h5" 
					dataSetPath="/MaxCoolingNull" powerUnit="kW"/>
    </Usage>
    <Grid id="GridConnection">
        <ElectricEnergyPrice fileName="UDK Heat Pump Scenario-2017-05.h5" 
					dataSetPath="/ECostFix" energyPriceUnit="ct/kWh"/>
        <ElectricEnergyRefund fileName="UDK Heat Pump Scenario-2017-05.h5" 
					dataSetPath="/ERefundFix" energyPriceUnit="ct/kWh"/>
    </Grid>
    <HeatBuffer id="HotWaterBuffer" initialThermalEnergyLevel="0.0" 
			energyUnit="kWh"/>
    <HeatPump id="HeatPump" isOnAtBegin="false" lastStartStopChangeInHours="0.5" 
			priceUnit="ct">
        <CoefficientOfPerformance fileName="UDK Heat Pump Scenario-2017-05.h5" 
					dataSetPath="/COP"/>
    </HeatPump>
</BuildingSituation>
\end{lstlisting}

In detail, usage profiles are stored in HDF5 files (with file extension ``{\tt .h5}''),  which are referenced in the 
respective XML elements (e.g., {\tt <ElectricPowerUsage>}). The name of the file containing the operation schedules 
must be entered as the value of the attribute {\tt fileNameHDF5} in the XML root element {\tt <BuildingSituation>}.
After processing the models the optimized schedule of the heat pump is stored in this file. 

\begin{figure}
  \begin{center}
		\vspace{-4mm}
		\includegraphics[width=1.0\textwidth]{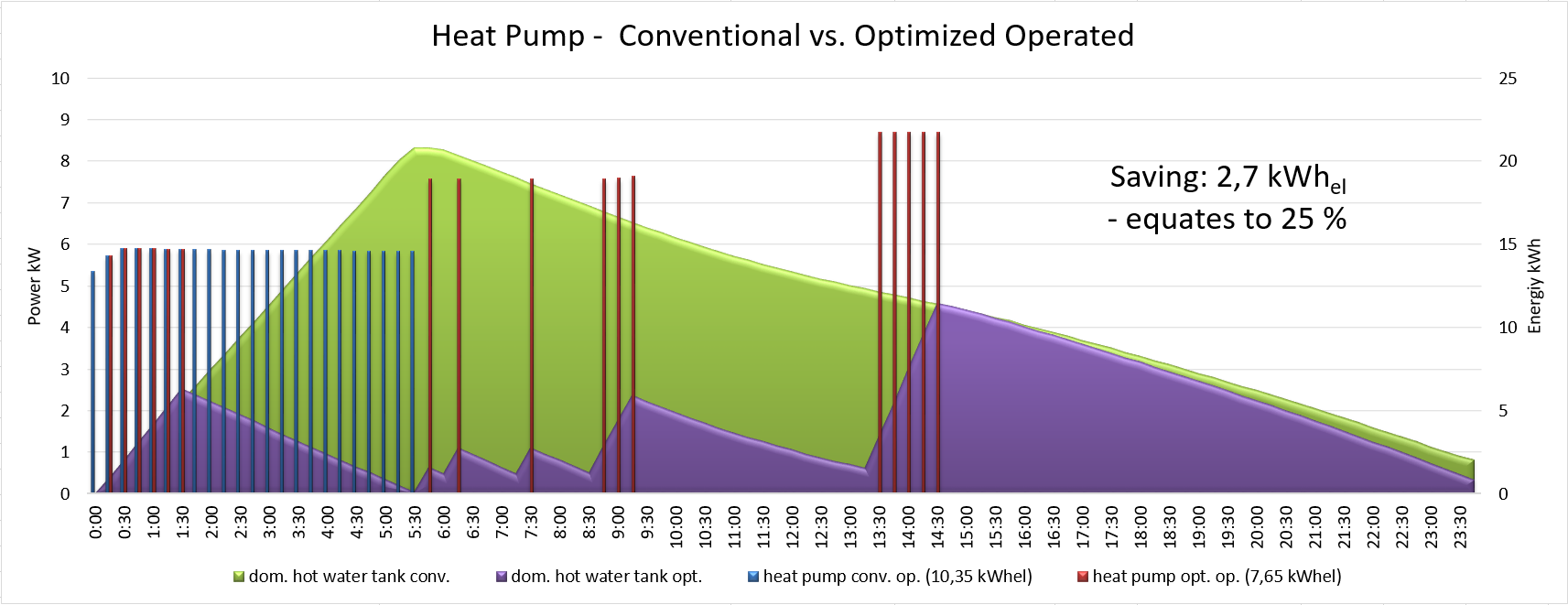}
		\caption{Heat pump conventional operation versus optimized operation.}
		\label{fig:heatPumpSample}
		\vspace{-8mm}
	\end{center}
\end{figure}

The resulting cost-minimized schedule of the operation of the building energy system 
(cf. Figure~\ref{fig:heatPumpSample}) shows that, in contrast to charging the thermal storage tank (TST) during the 
night hours (blue), an extensive operation of the heat pump at low outside temperatures and low COP can be avoided, 
if the heat demand and COP is known in advance and the heat pump is operated accordingly (red). 
Covering the currently predicted heat demand only enables a saving of 25~\%  electric energy, provided 
that the forecasts correspond to reality. Due to the general uncertainty of forecasts for energy supply and demand 
it is strongly recommended for practical applications to use some energy reservoirs, e.g., TSTs or batteries, and
parts of their charging/discharging capacities as buffers to balance deviations.  

\section{Conclusion and Future Work}

In this paper a MIP-based approach is presented to model and optimize the operation of building energy systems.
In detail the modeling of the energetic behavior of fcCHPs is presented and it is shown how sub-models of 
different energy component can be combined reflecting their integration into building energy systems. By example 
it is shown that the approach can be applied successfully. However, a comprehensive analysis on the accuracy and
adequacy of the MIP models and the impact of imperfect forecasts for energy demand/production and energy prices has
the be performed in the future, maybe in a follow-up research and development project. 

\bibliographystyle{plain}

\bibliography{localBib}

\begin{thebibliography}{10}

\bibitem{beasleyNotesSeparableProgramming}
John~E. Beasley.
\newblock {{OR Notes}} -- {{Separable}} programming.
\newblock http://people.brunel.ac.uk/\%7Emastjjb/jeb/or/sep.html.

\bibitem{bosmanPlanningProductionFleet2012}
M.~G.~C. Bosman, V.~Bakker, A.~Molderink, J.~L. Hurink, and G.~J.~M. Smit.
\newblock Planning the production of a fleet of domestic combined heat and
  power generators.
\newblock {\em European Journal of Operational Research}, 216(1):140--151,
  January 2012.

\bibitem{bozchaluiOptimalOperationCommercial2012}
Mohammad~Chehreghani Bozchalui and Ratnesh Sharma.
\newblock Optimal {{Operation}} of {{Commercial Building Microgrids Using
  Multi}}-{{Objective Optimization}} to {{Achieve Emissions}} and {{Efficiency
  Targets}}.
\newblock In {\em Power and {{Energy Society General Meeting}}, 2012 {{IEEE}}},
  pages 1--8, 2012.

\bibitem{brahmanOptimalElectricalThermal2015}
Faeze Brahman, Masoud Honarmand, and Shahram Jadid.
\newblock Optimal {{Electrical}} and {{Thermal Energy Management}} of a
  {{Residential Energy Hub}}, {{Integrating Demand Response}} and {{Energy
  Storage System}}.
\newblock {\em Energy and Buildings}, 90:65--75, 2015.

\bibitem{folkOverviewHDF5Technology2011}
Mike Folk, Gerd Heber, Quincey Koziol, Elena Pourmal, and Dana Robinson.
\newblock An overview of the {{HDF5}} technology suite and its applications.
\newblock In {\em Proceedings of the {{EDBT}}/{{ICDT}} 2011 {{Workshop}} on
  {{Array Databases}} - {{AD}} '11}, pages 36--47, {Uppsala, Sweden}, 2011.
  {ACM Press}.

\bibitem{forrestCBCUserGuide}
John Forrest.
\newblock {{CBC User Guide}}.
\newblock https://www.coin-or.org/Cbc/cbcuserguide.html.

\bibitem{thehdfgroupHDFHome}
The~HDF Group.
\newblock {{HDF Home}}.
\newblock https://www.hdfgroup.org/.

\bibitem{guOptimalConfigurationAnalysis2015}
Wei Gu, Yiyuan Tang, Shuyong Peng, Delin Wang, Wanxing Sheng, and Keyan Liu.
\newblock Optimal configuration and analysis of combined cooling, heating, and
  power microgrid with thermal storage tank under uncertainty.
\newblock {\em Journal of Renewable and Sustainable Energy}, 7(1):013104,
  January 2015.

\bibitem{guModelingPlanningOptimal2014}
Wei Gu, Zhi Wu, Rui Bo, Wei Liu, Gan Zhou, Wu~Chen, and Zaijun Wu.
\newblock Modeling, {{Planning}} and {{Optimal Energy Management}} of
  {{Combined Cooling}}, {{Heating}} and {{Power Microgrid}}: {{A Review}}.
\newblock {\em Electrical Power and Energy Systems}, 54:26--37, 2014.

\bibitem{ibmILOGCPLEXOptimization2019}
IBM.
\newblock {ILOG CPLEX Optimization Studio - Survey}.
\newblock https://www.ibm.com/products/ilog-cplex-optimization-studio, January
  2019.

\bibitem{jarvisaloModuleBasedFrameworkMultilanguage2009}
Matti J{\"a}rvisalo, Emilia Oikarinen, Tomi Janhunen, and Ilkka Niemel{\"a}.
\newblock A {{Module}}-{{Based Framework}} for {{Multi}}-language {{Constraint
  Modeling}}.
\newblock In Esra Erdem, Fangzhen Lin, and Torsten Schaub, editors, {\em Logic
  {{Programming}} and {{Nonmonotonic Reasoning}}}, Lecture {{Notes}} in
  {{Computer Science}}, pages 155--168. {Springer Berlin Heidelberg}, 2009.

\bibitem{steglichCMPLColiopMathematical2018}
Mike Steglich and Thomas Schleiff.
\newblock {{CMPL}}: {{Coliop Mathematical Programming Language}} - {{Version}}
  1.12 - {{March}} 2018, 2018.

\bibitem{w3cExtensibleMarkupLanguage}
W3C.
\newblock Extensible {{Markup Language}} ({{XML}}).
\newblock https://www.w3.org/XML/.

\bibitem{w3cXMLSchema}
W3C.
\newblock {{XML Schema}}.
\newblock https://www.w3.org/XML/Schema.

\end{thebibliography}

\end{document}